\documentclass[11pt,draftcls,onecolumn]{IEEEtran}
\usepackage{amsmath,graphicx}
\usepackage{enumerate}
\usepackage{amsbsy}
\usepackage{amssymb}
\usepackage{amsthm}
\usepackage{amscd}
\usepackage{subfigure}
\usepackage{color}
\usepackage{cite} 
\usepackage{dsfont}

\usepackage{balance}

\def\bSig{{\boldsymbol{\Sigma}}}

\def\bzero{{\boldsymbol{0}}}
\def\bone{{\boldsymbol{1}}}
\def\c1{{\textcircled{a}}}
\def\ba{{\boldsymbol{a}}}

\def\be{{\boldsymbol{e}}}

\def\bn{{\boldsymbol{n}}}

\def\bp{{\boldsymbol{p}}}

\def\bs{{\boldsymbol{s}}}

\def\bv{{\boldsymbol{v}}}
\def\bw{{\boldsymbol{w}}}
\def\bx{{\boldsymbol{x}}}
\def\by{{\boldsymbol{y}}}
\def\bz{{\boldsymbol{z}}}

\def\bB{{\boldsymbol{B}}}

\def\bD{{\boldsymbol{D}}}

\def\bH{{\boldsymbol{H}}}
\def\bI{{\boldsymbol{I}}}

\def\bM{{\boldsymbol{M}}}

\def\bQ{{\boldsymbol{Q}}}
\def\bR{{\boldsymbol{R}}}

\def\bV{{\boldsymbol{V}}}

\def\complexC{{\mathbb{C}}}

\def\integerZ{{\mathbb{Z}}}

\def\bzero{{\boldsymbol{0}}}
\def\bone{{\boldsymbol{1}}}

\newcommand{\mbf}[1]{\mathbf{#1}}

\DeclareMathOperator{\tr}{tr}

\title{Optimized Transmission for Parameter Estimation in Wireless Sensor Networks}

\author{Shahin~Khobahi, Mojtaba~Soltanalian, \emph{Member, IEEE}, Feng~Jiang,  and A.~Lee~Swindlehurst, \IEEEmembership{Fellow,~IEEE}
	\thanks{This work was supported in part by U.S. National Science Foundation Grants CCF-1704401, CCF-1703635, and ECCS-1809225. Parts of this work have been presented at the International Conference on Acoustics, Speech and Signal Processing (ICASSP), Calgary, Canada, April 2018 \cite{khobahi2018optimized}.
		\par S. Khobahi and M. Soltanalian are with the Department of Electrical and Computer Engineering, University of Illinois at Chicago, Chicago, IL 60607.  F. Jiang and A. L. Swindlehurst are with the Department of Electrical Engineering and Computer Science, University of California at Irvine, Irvine, CA 92697.
		}
}

\begin{document}

 \maketitle

\begin{abstract}
	
 A central problem in analog wireless sensor networks is to design the gain or phase-shifts of the sensor nodes (i.e. the relaying configuration) in order to achieve an accurate estimation of some parameter of interest at a fusion center, or more generally, at each node by employing a distributed parameter estimation scheme. In this paper, by using an over-parametrization of the original design problem, we devise a cyclic optimization approach that can handle tuning both gains and phase-shifts of the sensor nodes, even in intricate scenarios involving sensor selection or discrete phase-shifts. Each iteration of the proposed design framework consists of a combination of the Gram-Schmidt process and power method-like iterations, and as a result, enjoys a low computational cost. Along with formulating the design problem for a fusion center, we further present a consensus-based framework for decentralized
 estimation of deterministic parameters in a distributed network, which results in a similar sensor gain design problem. The numerical results confirm the computational advantage of the suggested approach in comparison with the state-of-the-art methods---an advantage that becomes more pronounced when the sensor network grows large.  
\end{abstract}
\section{Introduction} \label{sec:intro}
\IEEEPARstart{A}{nalog} vs. Digital? When it comes to wireless sensor networks (WSNs),  analog WSNs exhibit a significantly reduced level of distortion in parameter estimation compared to their digital counterparts \cite{Gastpar:2008}. Consequently, analog WSNs have been recently subject to extensive study---see, e.g., \cite{Cui:2007,Gastpar:2008, Smith:2009, Banavar:2012, banavar2010estimation, jiang2013estimation, feng2014power, feng2014massive}, and the references therein. 

The task of collecting an \emph{estimated} (or recovered) \emph{parameter} (or signal) from measurements of the sensor network is usually performed in a centralized manner, i.e. at a fusion center (FC). To further improve the estimation or detection performance in WSNs, the FC can be configured with multiple antennas \cite{Banavar:2012, jiang2013estimation, feng2014massive}. Several measurement relaying strategies have been proposed, including amplify-and-forward \cite{Smith:2009, Banavar:2012}, and phase-shift-and-forward schemes \cite{jiang2013estimation}.
It was shown that the transmission gain or phase-shift at the sensor nodes can be optimized in order to considerably reduce the estimation error at the FC. On the other hand, due to the fact that in a WSN the sensors are usually located in geographically different positions, the task of parameter estimation using a WSN requires the development of local signal processing techniques as well as developing inter-sensor communication strategies to further facilitate the estimation process. In addition, due to the limited bandwidth, cost, and energy budget available in WSNs, one should also consider the design of efficient compression techniques for local observations of each node so that it allows for low rate communications for node-to-node and node-to-FC transmissions. Hence, it is of importance to develop distributed estimation algorithms that allow for a low rate, yet optimized, communication strategy for both centralized and decentralized parameter estimation. To this end, we attempt in this work to provide a unifying optimization framework in the context of sensor transmission gain design for both centralized and decentralized parameter estimation in a WSN, and also propose a novel compression and diffusion strategy for inter-sensor communications which relies not only on the quality of the sensor observations but also the quality of the communication channel and observations of the neighboring nodes as well. 

There exists an extensive literature on the routing strategy optimization and sensor selection schemes for distributed networks to increase the efficacy of the distributed system while maintaining the accuracy of the estimation framework for both centralized or decentralized scenarios. The approaches used for addressing the design of energy-efficient routing and relaying schemes for such systems are based on convex and non-convex optimization techniques \cite{yang2009energy, saleem2014empirical, mansourkiaie2015optimal}. Researchers have looked at these problems from different perspectives, e.g., reducing communication cost \cite{4403230, bajwa2006compressive,4305386}, joint optimization of the routing and power allocating scheme \cite{lin2007asymptotically,liu2010joint,jakobsen2010dehar}, developing opportunistic-based routing protocols \cite{hsu2015delay,rahman2017eecor,vieira2012performance}, joint optimization of resources and routing in a distributed manner \cite{tang2013routing}, among others; interested reader may consult \cite{ulukus2015energy, sarkar2016routing} and the references therein for further details. In an effort to extend the lifetime of the network, several sensor selection algorithms were also proposed and studied in \cite{jiang2013estimation}. Furthermore, the authors in \cite{shah2013joint} have considered the problem of energy efficient distributed parameter estimation in a WSN and proposed a Fixed-Tree Relaxation-Based Algorithm (FTRA) in conjunction with a computationally efficient iterative distributed algorithm to jointly optimize the sensor selection and routing scheme in a WSN. In such scenarios, \cite{shah2013joint} shows that the well-known Shortest-Path Tree routing scheme is not optimal if one wishes to consider both the total communication cost and the estimation accuracy, while the authors provide a method with better trade-off when both of the criteria must be taken into account. In \cite{pereira2018parameter}, the authors considered the problem of distributed estimation of a vector-valued parameter in a WSN with faulty transducers and proposed a diffusion-averaging distributed Expectation-Maximization (EM) algorithm in order to perform the estimation task. Most notably, a class of distributed (but centralized) linear estimators based on reduced-dimensionality observations were developed in \cite{schizas2007distributed} to tackle the problem of estimating stationary random signals, and in particular, block coordinate descent based iterations to handle the estimation task for correlated sensor observations were proposed.

In the centralized parameter estimation scenario, a central node (i.e., the fusion center) collects the data from the sensor nodes in the network to perform the task of parameter estimation. Then, the FC applies an inference algorithm (e.g., the maximum likelihood estimation method) on the raw data to obtain the final estimate. However, if the data transmission to the FC is costly, the data processing and power limitations can be alleviated via employing a decentralized/distributed communication and estimation model in which each node performs the task of estimation \emph{locally} while restricting the data exchange between neighboring nodes. Another approach is to use low-resolution sampling techniques for addressing the power and data-rate limitations \cite{kong2018nonlinear,jedda2017massive,8645383,mo2018mimo,khobahi2018deep,stein2018one}. Early works in the context of analog estimation include the study of algorithms for data fusion in both centralized and decentralized scenarios. For instance, the authors in \cite{xiao05} have proposed an average consensus-based decentralized estimation scheme for a network with both fixed and time-varying topologies. In some recent efforts to achieve minimum estimation error, analog amplify-and-forward and phase-shift-and-forward transmission schemes for signal transmission from sensor to fusion center (FC) have been proposed in \cite{Cui:2007},\cite{Smith:2009}, \cite{Banavar:2012}, \cite{tsinos2018efficient}, and \cite{jiang2013estimation}, where the sensor gain optimization is usually subject to a total power constraint. Moreover, a distributed parameter estimation algorithm based on alternating direction method of multipliers (ADMM \cite{boyd11}) has been proposed in \cite{schizas08} and \cite{schizas08-2}. In addition, an ADMM-based method for phase-shift-and-forward (i.e., unit-modulus beamforming) wireless sensor networks aiming at estimating a deterministic parameter from noisy sensor measurements in a centralized manner has been proposed in \cite{8125772}. In particular, the authors of \cite{8125772} formulate the problem of centralized parameter estimation as a uni-modular quadratic program (UQP) and tackle it using the ADMM technique. Furthermore, the authors in \cite{6882386} have considered the scenario of detection of a zero-mean
Gaussian signal in a centralized manner and propose a convex sensor gain optimization framework by minimizing the ratio of the log probability of detection and log probability of false alarm under a fixed-norm gain constraint. Nevertheless, our work differs from \cite{8125772} and \cite{6882386} due to the fact that our proposed method can handle several practical scenarios (see Sec. III(A)-(B)) including, e.g., both the unit-modulus beamforming scenario of \cite{8125772} and fixed-norm gain design of \cite{6882386} in \emph{centralized} and \emph{decentralized} system architectures, and even more. For a more general overview of beamforming and sensor gain design techniques, the reader is referred to \cite{gershman2010convex} and \cite{haro2014energy}.
\vspace{5pt}
 \emph{Contributions:}
 In this work, we first formulate the problem of parameter estimation for both cases of centralized and decentralized data fusion models. Then, we derive the asymptotic variance of the estimation for both cases and we propose an \emph{efficient} framework that can deal with tuning both gains and phase-shifts of the sensors for an optimized forwarding of the observed signal for node-to-node and node-to-FC communication purposes, which effectively minimizes the final error variance of the estimation, facilitating a better estimation accuracy for the parameter in both the decentralized and centralized scenarios. Furthermore, we propose a novel data compression and communication strategy for the decentralized estimation scenario, and further show that the centralized and decentralized estimation frameworks can be viewed under a single unified optimization model. In addition, the corresponding mean-squared-error (MSE) performance of the proposed estimation techniques for both scenarios are derived in terms of the sensor gain vectors. In particular,
 
 \begin{itemize}
 	\item The proposed algorithm can deal with the optimization of the complex gains of the sensors (i.e. both phase-shifts and transmit gains) for both \emph{centralized} and \emph{decentralized} parameter estimation models. In addition, our method offers an extremely low computational complexity compared to state-of-the-art methods. In particular, in the centralized parameter estimation scenario, our proposed optimization method demonstrates far better estimation accuracy compared to other methods. In addition, it is superior in computational performance for large-scale distributed systems, compared to the state-of-the-art SDP-based approaches.
 	\item The proposed approach can be used for various types of sensor constraints including e.g., phase-shift only and sensor selection cases.
 	\item In the phase-shift only case, we propose a simpler alternative to our general framework.
 \end{itemize}
 

\vspace{5pt}
\emph{Organization of the Paper:}
The rest of the paper is organized
as follows. In Section II, we give the general problem formulation of the both decentralized and centralized parameter estimation schemes with their associated data fusion algorithm. In Section III, we propose our efficient sensor gain optimization technique which effectively minimizes the variance of the estimation methods for both the cases of decentralized and centralized parameter estimation. In Section IV, we thoroughly investigate the performance of our proposed sensor gain optimization for different scenarios and we compare our algorithm with several state-of-the-art methods. Finally, Section V provides a summary that concludes the paper.

\vspace{5pt}
\emph{Notation:} We use bold lowercase letters for vectors and bold uppercase letters for matrices. $(\cdot)^T$, and $(\cdot)^H$ denote the vector/matrix transpose, and the Hermitian transpose, respectively. $\bone$ and $\bzero$ are the all-one and all-zero vectors/matrices.  $\| \bx \|_n$ or the $l_n$-norm of the vector $\bx$ is defined as $\left( \sum_k |\bx(k)|^n \right)^\frac{1}{n}$ where  $\{ \bx(k) \}$ are the entries of $\bx$. The symbol $\odot$ stands for the Hadamard matrix product. $\mathbf{Diag}(\cdot)$ denotes the diagonal matrix formed by the entries of the vector argument, and $\mathbf{blkdiag}(\cdot)$ returns a block diagonal matrix with matrices on its diagonal. $\integerZ_Q$ denotes the set $\{ 0,1, \cdots, Q-1 \}$. We represent the topology of the WSN by an undirected and connected graph $\mathcal{G}=(\mathcal{E},\mathcal{V})$, consisting of a finite set of vertices $\mathcal{V}=\{1,\dots,n\}$  (also called \textit{nodes}), and a set of edges $\mathcal{E}\subseteq\{\{i,j\}:~i,j\in\mathcal{V}\}$. We denote the edge between node $i$ and $j$ as $\{i,j\}$, which indicates a bidirectional communication between the nodes $i$ and~$j$. We further assume that the sensor connections in $\mathcal{G}$ are time-invariant and the transmissions are always successful.
We define the set of neighbors of node $i$ including itself as $\mathcal{N}_i\triangleq \{j\in \mathcal{V} : ~ \{i,j\} \in \mathcal{E}\}$. The \textit{degree} of the $i$th node is given by $d_{i} = |\mathcal{N}_{i}|$.

\section{System and Fusion Model}
In this section, we consider two different data exchange scenarios in a multi-agent network (e.g., a WSN) where the ultimate goal is to achieve a maximum likelihood (ML) estimation of some observed parameter. We will refer to the two scenarios as centralized and decentralized data exchange schemes. We further present a consensus-based framework for decentralized estimation of deterministic parameters in wireless sensor networks and further show that the error variance in this case converges to that of the global maximum likelihood estimate of the parameter when a central node has access to all information available in the network. Next, we propose an efficient sensor gain optimization technique to minimize the overall error variance of the estimate derived in both centralized and distributed frameworks.

\subsection{Problem Formulation: Decentralized Estimation}
\begin{figure*}[t]
	\centering
	
	\includegraphics[width=\linewidth]{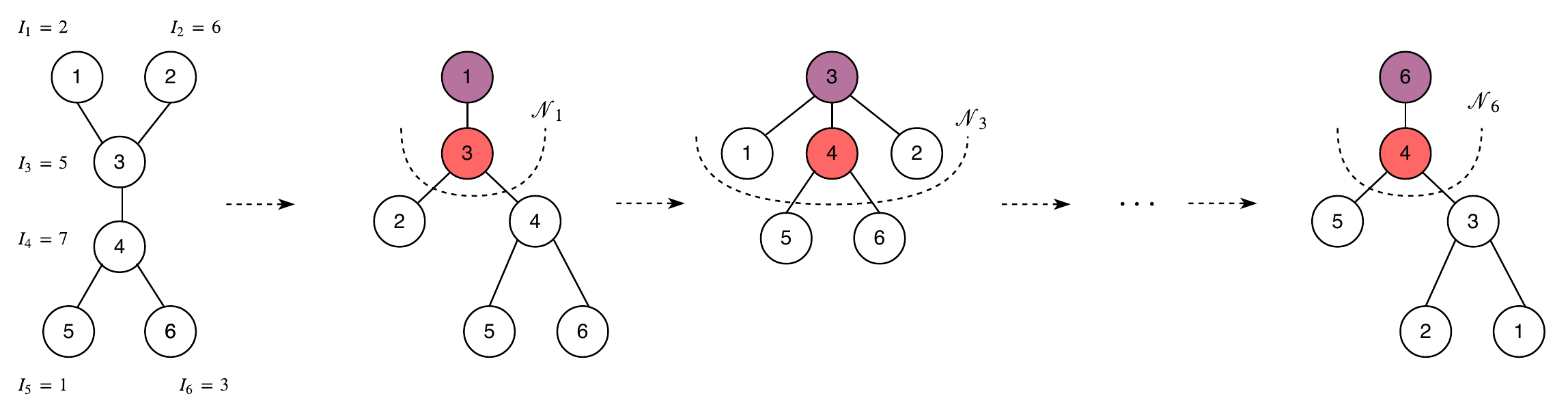}
	\caption{An intuitive illustration of the proposed data compression and diffusion strategy for a graph with $N=6$ nodes. Purple nodes represent the parent nodes at each data compression phase, while the red ones show the node that is assigned to carry the information received from the corresponding parent node. Each node starts with an initialization of the information variable $I_i(0)$ for $i=1,\dots,N$, and transmits this variable to its immediate neighbors. Next, each node (denoted by the color purple, considered as a parent node in its neighborhood) assigns the node with the highest information value (red node) to incorporate the parent node's observation into its diffusion process.}\label{f1}
	\label{fig:toy}
\end{figure*}
We present a distributed consensus-based estimation framework for our problem and further derive the error variance formulation for this case, which will be utilized as a metric to optimize the transmission gains for our amplify and forward node-to-node data exchange protocol. We consider a network with $N$ single-antenna agents (nodes) each of which observes an unknown deterministic parameter $\theta\in\mathds{C}$, according to the following linear observation model:
\begin{align}
\label{eqs:1}
z_i = \theta+v_i,
\end{align}
where $i$ denotes the sensor index, and $v_i\sim\mathcal{CN}(0,\sigma_{v,i}^2)$ is additive Gaussian observation noise. We further assume that the observation noise is independent from one node to another, and that the channel state information (CSI) of the network is available at the nodes (at least for neighbors). 
\par Our distributed data exchange protocol operates as follows. The $i$th agent amplifies its observation $z_i$ with an adjustable complex gain $a_i\in\mathds{C}$, and then, transmits this amplified observation to its immediate neighbors (i.e., $k\in\mathcal{N}_i$). Hence, the received vector at a generic node $i$ from its neighbor node $k$ can be modeled as:
\begin{equation}
\label{eqs:2}
y_{i,k} = h_{i,k} a_k z_k + n_{i}.\quad \text{for } k\in\mathcal{N}_i
\end{equation}
where ${h_{i,k}}\in\mathds{C}$ denotes the complex gain of the channel between nodes $k$ and $i$, and $n_{i}\sim\mathcal{CN}(0,\sigma^2_{n})$ is the zero-mean Gaussian communication noise which is assumed to be uncorrelated from one transmission to another. Let $S^{i}=\{s^i_1,\dots,s^i_{|\mathcal{N}_i|}\}$ denote the ordered sequence of all nodes neighboring the $i$-th node. This ordered set allows for a concise description of the estimation and system model in the rest of the paper. Considering the generic node $i$ to be a data sink point in the network, the vector of all observations of its neighboring nodes prior to amplification can be modeled as,
\begin{equation}
\label{eqs:3}
	\boldsymbol{z}_{i} = \theta\mbf{1} + \bv_{i},
\end{equation}
where $\boldsymbol{z}_{i} = [z_{s_1^i},\dots,z_{s^i_{|\mathcal{N}_i|}}]^T$, and $\bv_{i}=[v_{s_1^i},\dots,v_{s^i_{|\mathcal{N}_i|}}]^T$ denotes the vector of observation noise for all nodes in the neighborhood of the $i$-th node (i.e., $k\in\mathcal{N}_i$), for which the covariance of the aggregated noise vector is given by
$$\bR_{\bv_i}=\mathds{E}\{\bv_i\bv_i^H\}=\mathbf{Diag}\left(\sigma^2_{v,s_1^i},\dots,  \sigma^2_{v,s_{|\mathcal{N}_i|}^i}   \right).$$
Consequently, the received vector of the \textit{amplified} signal at node $i$ from its neighbors can be expressed as:
\begin{equation}
\label{eqs:4}
	\by_i = \mbf{H}_i \mbf{D}_i \bz_i + \bn_i = 
	\mbf{H}_i \ba_i \theta + \underbrace{ \mbf{H}_i \mbf{D}_i \bv_i + \bn_i}_{\triangleq\bw_i},
\end{equation}
where $\ba_i=[a_{s_1^i} , \dots , a_{s^i_{|\mathcal{N}_i|}}]^T$ denotes the \emph{complex sensor gains} to be optimized, $\mbf{D}_i = \mathbf{Diag}(\ba_i)$, $\by_i=[y_{i,s_1^i} , \dots , y_{i,s^i_{|\mathcal{N}_i|}}]^T$ whose elements are defined in \eqref{eqs:2}, $\mbf{H}_i=\mathbf{Diag}\left(h_{i,s_{1}^i} ,\dots, h_{i,s_{|\mathcal{N}_i|}^i}\right)$, and $\bn_i$ is complex Gaussian communication noise vector at the sink node $i$, with covariance matrix $\bR_{\bn_i} = \mathds{E}\{\bn_i\bn_i^H\} = \sigma_n^2\mbf{I}_{|\mathcal{N}_i|}$
. In addition, according to \eqref{eqs:4}, we define $\bw_i \triangleq \mbf{H}_i \mbf{D}_i\bv_i + \bn_i$, as the \emph{combined} noise term for the communication noise and the amplified measurement noise. Clearly, the combined noise term $\bw_i$ follows a zero-mean Gaussian distribution with covariance matrix
\begin{equation}
\label{eqs:5}
	\mbf{R}_{\bw_i} = \mathds{E}\{\bw_i\bw_i^H\} = \mbf{H}_i \mbf{D}_i \mbf{R}_{\bv_i}\mbf{D}_i^H\mbf{H}_i^H + \mbf{R}_{\bn_i}.
\end{equation}
\subsection{Local Estimation Model}
Let $\hat{\theta}_i^{(t)}$ denote the local estimate of the unknown parameter at node $i$ after $t$ rounds of data exchange in the network. Henceforth, according to \eqref{eqs:4} and after the first round of node-to-node communication (data exchange via the amplify-and-forward scheme) in the network, each node can compute the initial maximum likelihood estimate of the parameter based on the received information from its neighboring nodes as
\begin{equation}
\label{eqs:6}
	\hat{\theta}_i^{(1)} = \left(\ba_i^H\mbf{H}_i^H\mbf{R}_{\bw_i}^{-1}\mbf{H}_i\ba_i\right)^{-1}
	\ba_i^H \mbf{H}_i^H\mbf{R}_{\bw_i}^{-1}\by_i,
\end{equation}
where the error variance of the above estimate is given by
\begin{equation}
\label{eqs:7}
\text{Var}\left(\hat{\theta}_i^{(1)}\right) = \left(\ba_i^H\mbf{H}_i^H\mbf{R}_{\bw_i}^{-1}\mbf{H}_i\ba_i\right)^{-1}.
\end{equation}
\par A drawback of such a distributed \emph{amplify-and-forward} scheme (governed by variable sensor gains) is that all nodes neighboring a specific node $i$ will receive the amplified noisy observation of that node, $y_{i,k}\; \text{for}\; k\in\mathcal{N}_i$, and further incorporate that single noisy observation into their estimation model. In such an approach, the aggregated global data of the entire network is correlated. In order to alleviate this problem, and further reduce the diffusion of redundant information in the network, we propose the following data compression strategy which results in a more efficient node-to-node communication in that it allows the diffusion of the most useful (less contaminated) data in the network.
\subsection{Data Compression and Diffusion Strategy}
%
Each node starts an initial local \emph{information} variable $I_i(t)$ for $i\in\mathcal{V}$, according to the following model (where $t$ denotes the discrete time index):
\begin{equation}
\label{eqs:8}
	I_i(0) = \ba_i^H\mbf{H}_i^H\mbf{R}_{\bw_i}^{-1}\mbf{H}_i\ba_i.
\end{equation}
Note that the above initialization scheme requires each node to have the (partial) channel state information, the transmission gains, and the second-order noise statistics of its neighboring nodes, which we assume are available at all nodes. Each node then transmits its initial local information variable $I_i(0)$ to its immediate neighbors. Furthermore, note that $I_i(0)$ can be seen as a measure of information in that the inverse of $I_i(0)$ provides the local error variance of the maximum likelihood estimate (MLE) of the unknown parameter at each node. After this round of information exchange, each node (which also can be seen as a \emph{parent} node in its local neighborhood) will select the node from its neighbors with the highest information value and only the selected node will retain/use the received data from the parent node; all other nodes will discard their associated received signal for that particular parent node. In the case that more than one node has the highest information value, the parent node can choose either one for relaying purposes (e.g., by random assignment) as our diffusion strategy is based on one-hop information. In other words, assume that the $j$-th node has the highest information value among the $i$-th node's neighborhood. Then, all nodes $k\in\mathcal{N}_i\backslash\{j\}$ will discard $y_{i,k}$, except the $j$-th node.

In order to further clarify the above described data compression strategy, consider the toy graph example with $N=6$ nodes illustrated in Fig. \ref{fig:toy}. In this case, we assume that the initial information values of the nodes are given by the vector $I(0) ~=~[2,6,5,7,1,3]^T$, respectively. As it can be seen from the illustrated graph, node 1 has only one immediate neighbor which is node 3, and thus, it has the highest information value in $\mathcal{N}_1$. Therefore, node 1 assigns node 3 to carry the information received from node 1 in the graph. Similarly, consider node 3 as the parent node where $\mathcal{N}_3=\{1,2,4\}$ with corresponding information values of $[2,6,7]$ respectively. Consequently, node 4 now has the highest information value among the immediate neighbors of node 3 (the parent node), and thus, node 3 assigns node 4 for processing the information received from node 3 in $\mathcal{N}_3$, and all other nodes in the neighborhood of the parent node 3 will discard the received data from it, i.e., the nodes $\mathcal{N}_{3}\backslash\{4\} = \{1,2\}$. Eventually, in a similar manner, the nodes $\{3,5,6\}$ assign node $4$ as the carrier of their transmitted information, and the nodes $\{1,2,4\}$ choose node $3$ as the carrier of their information.

\par The above local estimation and data compression scheme can be described as follows. Let $\{\mbf{T}_i\}_{i=1}^{N}$ denote the \emph{row selection matrix} associated with the $i$-th node, which points to the rows of the vector $\by_i$ that are to be retained according to the above data compression strategy. Thus, the aggregated received data at each node and after applying the proposed compression strategy is given by
$$\by^{\prime}_i = \mbf{T}_i\by_i.$$
For instance, in the toy example illustrated in Fig \ref{fig:toy}, node 4 is assigned to retain the information received from the nodes $\{3,5,6\}$. Thus, its row selection matrix should be set as $\mbf{T}_4=\mbf{I}_3$, where $\mbf{I}_n$ denotes the identity matrix of dimension $n$. On the other hand, nodes $\{1,2,5,6\}$ have not been chosen by any node for information diffusion purposes, and therefore, their row selecting matrix is the all-zero square matrix of dimension $|\mathcal{N}_i|$ for $i\in\{3,5,6\}$, respectively.

The compressed \emph{global} observation vector collected from all nodes following the above described compression strategy can be modeled as follows:
\begin{equation}
\label{eqs:9}
\by = \mbf{H} \ba \theta + \mbf{HD}\bv + \mbf{G}\bn,
\end{equation}
where $\ba=[a_1,\dots,a_N]^T$ is the vector of complex gains to be optimized, $\mbf{D} = \mathbf{Diag}\left(\ba\right)$, $\bv=[v_1,\dots,v_N]^T$ denotes the vector of observation noise at all nodes whose covariance matrix is given by $\mbf{R}_\bv=\mathbf{Diag}(\sigma^2_{v,1},\dots,\sigma^2_{v,N})$, $\mbf{G}=\mathbf{blkdiag}\left(\{\mbf{T}_i\}_{i=1}^N\right)$, and $\mbf{H} = [\mbf{T}_1\boldsymbol\Omega_1,\dots,\mbf{T}_N\boldsymbol\Omega_N]^T$ where the matrix $\boldsymbol{\Omega}_i$ is a $|\mathcal{N}_i|\times N$ matrix whose elements are defined as follows:
\begin{equation}
	[\boldsymbol{\Omega}_i]_{m,n} = 
	  \begin{cases}
		h_{i,n} & \text{if $n\in S^i$ and $n = s_m^i$}, \\
		0 & \text{otherwise}.
	\end{cases}
\end{equation}
Moreover, we define the global combined noise term in \eqref{eqs:9} as
\begin{equation}
	\bw = \mbf{HD}\bv + \mbf{G}\bn,
\end{equation}
which is zero-mean Gaussian noise with covariance matrix
\begin{equation}
\label{eqs:11}
\mbf{R}_\bw = \mathds{E}\{\bw\bw^H\} = \mbf{HDR_{\bv}}\mbf{D}^H\mbf{H}^H + \mbf{M},
\end{equation}
where $\mbf{M} = \sigma^2_n\mbf{I}_{M}$, and $M = 2|\mathcal{E}|-r$, where $r$ represents the total number of discarded communications emanating from the proposed data compression strategy.
\par Following the global data model of \eqref{eqs:9}, the maximum likelihood estimate of the unknown parameter can thus be expressed as:
\begin{align}
\label{eqs:13}
\hat{\theta}_{ML} &= \left(\ba^H\mbf{H}^H\mbf{R}_\bw^{-1} \mbf{H} \ba\right)^{-1}\ba^H\mbf{H}^H\mbf{R}_{\bw}^{-1}\by\\
	&=\left(\sum_{i=1}^N  \ba_i^H\mbf{H}_i^H\mbf{R}_{\bw_i}^{-1}\mbf{H}_i\ba_i  \right)^{-1}\sum_{i=1}^{N}\ba_i^H\mbf{H}_i^H\mbf{R}_{\bw_i}^{-1}\by_i^{\prime},
	\label{eqs:14}
 \end{align}
where, for the sake of simplicity, we assume that in \eqref{eqs:14}, the local sensor gain vector $\ba_i$, the diagonal channel matrix $\mbf{H}_i$, and diagonal noise covariance matrix $\mbf{R}_{\bw_i}$ only contain the values associated with the nodes whose information has not been discarded as a result of suggested data compression strategy at each node. Furthermore, note that the maximum likelihood estimate $\hat{\theta}_{ML}$ is unbiased (i.e., $\mathds{E}\{\hat{\theta}_{ML}\}=\theta$) with variance,
\begin{align}
\nonumber
\text{Var}(\hat{\theta}_{ML}) &= \left(\ba^H\mbf{H}^H\mbf{R}_\bw^{-1} \mbf{H} \ba\right)^{-1}\\
											&=\left(\sum_{i=1}^N  \ba_i^H\mbf{H}_i^H\mbf{R}_{\bw_i}^{-1}\mbf{H}_i\ba_i  \right)^{-1}.
											\label{eqs:15a}
\end{align}
\par\emph{Remark 1:} The maximum likelihood estimation of the parameter presented in \eqref{eqs:13} can be split into the summation of $N$ terms of the form \eqref{eqs:14} as a result of our data compression and diffusion strategy. Namely, the proposed compression technique diffuses the information of each node in such a way that each amplified observation only appears \emph{once} in the network. Indeed, there exists only \emph{one} node in the neighborhood of a generic node $i$ that incorporates the amplified observation of the $i$-th node into its estimation model. Therefore, the amplified observations are uncorrelated across the whole network resulting in the expressions \eqref{eqs:13} and \eqref{eqs:14}. $\blacksquare$
\par Our goal is to facilitate computing the \emph{global} maximum likelihood estimate of the parameter presented in \eqref{eqs:13} and \eqref{eqs:14} in a distributed manner. Namely, we employ an average-consensus scheme based on the alternating direction method of multipliers (ADMM) \cite{boyd11} enabling us to asymptotically converge to the global MLE of the parameter via local computations while allowing only communication between the nodes and their immediate neighbors. Finally, we derive the error variance for the distributed estimation algorithm which will be used as a metric for optimizing the complex sensor gains.
\subsection{ADMM-Aided Distributed ML Estimation}
We begin this part by describing the general consensus problem in a multi-agent system, and then, use an ADMM-based average consensus algorithm to solve the maximum likelihood estimation of \eqref{eqs:14} in a decentralized and distributed manner.
\vspace{5pt}
\par \emph{\underline{Consensus}:}
Consider a group of agents $i\in\{1,\dots,N\}$ each of which has access to a local variable $x_i$ associated with its initial observation, and let $\bx=[x_1,\dots,x_n]^T$ denote the stacked column vector of the variables for all agents. The aim of the distributed average-consensus algorithms is to find the average of the local variables, e.g., $x_{\text{avg}} =\frac{1}{N} \mbf{1}^T\bx$, in a distributed manner and via restricting collaborations to be between adjacent agents. In addition, finding the average value of the local variables can be recast as the unique solution of the following unconstrained least-squares program:
\begin{align}
\label{eqs:15}
x_{\text{avg}} = \underset{y}{\text{argmin}}\;\frac{1}{2}\sum_{i=1}^{N}(y - x_i)^2 = \underset{y}{\text{argmin}}\;\frac{1}{2}||y\mbf{1}-\bx||_2^2.
\end{align}

The goal is next to compute \eqref{eqs:15} in a decentralized manner by allowing only local data-exchanges in each neighborhood. In order to do so, the minimization of \eqref{eqs:15} can be further reformulated as a global consensus probem via utilizing the underlying network (graph) connectivity structure. Namely, we first decouple the unconstrained program of \eqref{eqs:15} by introducing local copies of the global variable $y$ at each node, and then, enforcing the local copies to be equal across the network. This reformulation of \eqref{eqs:15} can be expressed as follows:
\begin{align}
\nonumber
&\underset{\{y_i\},\{c_{i,j}\}}{\text{minimize}}\quad\sum_{i=1}^{N} \frac{1}{2}(y_i - x_i)^2\\
& s.t.\quad y_i = c_{i,j},\;\;y_j = c_{i,j}, \forall(i,j)\in\mathcal{E},\label{eqs:16}
\end{align} 
where $y_i$ is the $i$-th node's local copy of the global variable $y$, and $c_{i,j}$ are auxiliary variables ensuring consensus between the neighboring nodes. Note that for a connected graph where there exists at least one path (a chain of edges) between any two nodes in the network, the two problems in \eqref{eqs:15} and \eqref{eqs:16} are equivalent. It is noteworthy to mention that problem \eqref{eqs:15} is centralized in that it requires all of the information (e.g., $\{x_i\}_{i=1}^{N}$) in the network to find the optimal solution $x_{\text{avg}}$. On the other hand, the new (equivalent) program in \eqref{eqs:16} requires each node $i$ to find the local variable $y_i$ that is optimal for the overall objective function $R\triangleq(1/2)\sum_i(y_i-x_i)^2$, without having global knowledge of the observations at other nodes.
\par Herein, we use the alternating direction method of multipliers (ADMM) to efficiently solve \eqref{eqs:16} in a distributed manner. In particular, the following ADMM \emph{update equations} were derived in \cite{shi14} to efficiently solve \eqref{eqs:16} and to eventually achieve an average-consensus in the network:
\begin{align}
\label{eqs:17}
&y_i^{k+1} = \frac{1}{1+2\rho|\mathcal{N}_i|}\Big(\rho |\mathcal{N}_i|y_i^{k} + \rho\sum_{j\in\mathcal{N}_i} y_j^k - \lambda_i^k+x_i  \Big),\\
&\lambda_i^{k+1} = \lambda_i^k + \rho\Big( |\mathcal{N}_i|y_i^{k+1} - \sum_{j\in\mathcal{N}_i} y_j^{k+1}  \Big),
\label{eqs:18}
\end{align}
where $k$ denotes the iteration number, $y^{k+1}_i$ is the local copy of the global variable at node $i$ (which will eventually converge to the average value of the initial observations $x_{\text{avg}} = (1/n)\sum_{i=1}^n x_i$), $x_i$ is the initial observation of node $i$, and $\rho>0$ is an arbitrary constant.  As it can be seen from above update equations, the updates of each node only depend on the \emph{local information}, and the algorithm is hence fully distributed. Next, we use this ADMM-based distributed average-consensus scheme to determine the ML estimate of the unknown parameter, i.e., solving \eqref{eqs:14} using \eqref{eqs:17}-\eqref{eqs:18}.
\vspace{5pt}
\par \emph{\underline{Distributed maximum likelihood estimation}:} In order to use the average-consensus algorithm to compute the MLE of the parameter in a distributed and decentralize manner, we follow a similar approach to the one proposed in \cite{xiao05}, and first assign the following initial variables to each node, which will be further used as the local observation of each node in the average-consensus algorithm:
\begin{align}
	\label{eqs:19}
	\text{\emph{Information Value}:}\quad &I_i(0)\triangleq\ba_i^H\mbf{H}_i^H\mbf{R}_{\bw_i}^{-1}\mbf{H}_i\ba_i,\\
	\text{\emph{State Information Value}:}\quad &P_i(0)\triangleq\ba_i^H\mbf{H}_i^H\mbf{R}_{\bw_i}^{-1}\by_i^{\prime}.
	\label{eqs:20}
\end{align}
Note that the MLE of $\theta$ in \eqref{eqs:14} is comprised of two terms: the inverse of the summation of the information values at each node $(\sum_i I_i(0))^{-1}$, multiplied by the summation of state information values $\sum_i P_i(0)$.
Henceforth, each node can (asymptotically) compute the \emph{global} ML estimate of the parameter $\theta$ defined in \eqref{eqs:14} by separately applying the distributed average consensus in \eqref{eqs:17} and \eqref{eqs:18} on the local variables, $I_i(0)$ and $P_i(0)$. More precisely, each node updates its information value and the state information value according to \eqref{eqs:17}-\eqref{eqs:18} (by substituting $x_i$ in \eqref{eqs:17} with $I_i(0)$ and $P_i(0)$), and will obtain a local estimate of the parameter of interest at each iteration by computing
\begin{equation}
\label{eqs:21}
\hat{\theta}^{i}(k) = I_i(k)^{-1}P_i(k).
\end{equation}

The asymptotic behaviour of $I_i(k)$ and $P_i(k)$ in the averge-consensus algorithm can be calculated as follows,
\begin{align}
\label{eqs:22}
I_c \triangleq &\;\;\underset{{k\rightarrow\infty}}{\text{lim}}\; I_i(k) = \frac{1}{N} \sum_{i=1}^{N}\ba_i^H\mbf{H}_i^H\mbf{R}_{\bw_i}^{-1} \mbf{H}_i\ba_i,\\
P_c \triangleq &\;\;\underset{{k\rightarrow\infty}}{\text{lim}}\; I_i(k) = \frac{1}{N} \sum_{i=1}^{N}\ba_i^H\mbf{H}_i^H\mbf{R}_{\bw_i}^{-1} \by_i^{\prime}.
\label{eqs:23}
\end{align}
Therefore, the local ML estimate of the unknown parameter at each node (i.e., $\hat{\theta}^{i}(k)$) will eventually converge to that of the global MLE of $\theta$ defined in \eqref{eqs:14}, i.e.,
\begin{equation}
\label{eqs:24}
\hat{\theta}^{i}_{ML} = I_c^{-1}P_c = \frac{\sum_{i=1}^{N}\ba_i^H\mbf{H}_i^H\mbf{R}_{\bw_i}^{-1} \by_i^{\prime}}{\sum_{i=1}^{N}\ba_i^H\mbf{H}_i^H\mbf{R}_{\bw_i}^{-1} \mbf{H}_i\ba_i}.
\end{equation}
Note that the average scaling factor $\frac{1}{N}$ is eliminated in \eqref{eqs:21} and \eqref{eqs:24}, and therefore, it will not affect the ML estimate at each iteration. In addition, it can be easily shown that the variance of the estimate at each node converges to that of the global ML etimation variance in \eqref{eqs:15a}, i.e.,
\begin{align}
\nonumber
	\underset{{k\rightarrow\infty}}{\text{lim}}\text{Var}\left(\hat{\theta}_{ML}^{i}(k)\right) &=\left(\sum_{i=1}^N  \ba_i^H\mbf{H}_i^H\mbf{R}_{\bw_i}^{-1}\mbf{H}_i\ba_i  \right)^{-1}\\
			&= \left(\ba^H\mbf{H}^H\mbf{R}_\bw^{-1} \mbf{H} \ba\right)^{-1}
			\label{eqs:25}.
\end{align}
By further substituting \eqref{eqs:11} into \eqref{eqs:25}, we have the following asymptotic expression for the error variance at each node:
\begin{equation}
\label{eqs:26}
\text{Var}(\hat{\theta}_{ML}) = \left(\ba^H\mbf{H}^H(\mbf{HDR_{\bv}}\mbf{D}^H\mbf{H}^H + \mbf{M})^{-1} \mbf{H} \ba\right)^{-1},
\end{equation}
where $\text{Var}(\hat{\theta}_{ML})$ denotes the asymptotic estimation variance of each node after convergence (and after reaching a consensus). Note that the proposed data compression and diffusion strategy plays a vital role in decoupling the sensor observations throughout the network, and specifically, it paves the way for optimizing the sensor gains without dealing with correlated data. On the other hand, the decentralized ADMM algorithm described above together with the proposed compression technique enables us to derive a closed form expression for the asymptotic variance of the estimation error based on the sensor gains, while allowing the network to achieve a very fast consensus on the global ML estimate. In Section 3, we devise a low-cost cyclic optimization approach to design the complex gains at each node via optimizing \eqref{eqs:26}.
\subsection{Problem Formulation: Centralized Estimation}
We now assume that there exists a fusion center aggregating the information received from the nodes to perform the task of parameter estimation in a centralized manner. In this case, the derivation of the error variance is straightforward and resembles the same structure as the decentralized case given in \eqref{eqs:26}.
\par We consider a network of $N$ sensors that observe an unknown parameter, where a maximum likelihood (ML) estimate of the unknown parameter is formed at the FC with $M$ antennas. As indicated earlier, it was shown in \cite{Smith:2009,Banavar:2012}, and \cite{jiang2013estimation} that the parameter estimation performance at the FC can be significantly improved by a judicious design of the complex relaying gains of the sensors. The variance of  the ML estimate of the parameter is given by
\begin{equation} \label{eq:var}
\mbox{Var} \left(\hat{\theta}_{ML} \right)=\left({\ba}^{H}{\bH}^{H}\left({\bH \bD \bSig  \bD^H \bH }^{H}+\bM\right)^{-1}{\bH \ba }\right)^{-1}
\end{equation}
where $\theta$ is the parameter to be estimated, $\bH \in \complexC^{M \times N}$ denotes the channel matrix, $\bSig \in \complexC^{N \times N}$ is the covariance matrix of the sensor measurement noise, $\bM \in \complexC^{M \times M}$ denotes the covariance matrix of the noise at the FC (similar to the communication noise in the decentralized case), $\ba \in \complexC^{N}$ comprises the adjustable \emph{complex} gains of the sensors, and $\bD=\mathbf{Diag}(\ba)$. Moreover, we intentionally use the same notation for the centralized estimation problem as for the decentralized one to emphasize the fact that the two gain optimization problems boil down to the same formulation.

\section{Sensor Gain Optimization} \label{sec:opt}
Hereafter, we address the problem of designing the (possibly complex) sensor gains $\ba\in\mathds{C}^N$ in order to minimize the variance of both the consensus-based distributed estimate given in \eqref{eqs:26}, and the centralized estimate scenario defined in \eqref{eq:var}. As shown in the previous section, for the decentralized MLE, the variance of the estimation at each node asymptotically converges to that of the global ML estimate of the unknown parameter. Furthermore, for the centralized estimation case, the error variance follows the same structure as the decentralized case paving the way for proposing a \emph{general} gain optimization framework for both cases. It is worth mentioning that although we have considered the problem of estimating a deterministic source signal $\theta$ in \eqref{eqs:1}, our formulations are also valid for more complicated scenarios such as multi-dimensional correlated sources \cite{4378416}. In this paper, we assume that the source signal $\theta$ is deterministic; however, in the case that the parameter of interest is probabilistic in nature, one can make use of the available knowledge on the parameter of interest in the form of a prior distribution $\bp(\theta)$, and approach the problem through a Bayesian framework. In this case, given the sensors observation vector $\bz=[z_1,\dots,z_N]$, one can consider the maximum a posteriori (MAP) estimator framework and evaluate $\mbox{argmax}_{\theta}\; \bp(\theta|\bz)$ in a \emph{distributed} manner where the posterior distribution is obtained by utilizing the Bayes' rule, i.e., $\bp(\theta|\bz) = \bp(\bz|\theta)\bp(\theta)/\bp(\bz)$ (sensor gains vector must be further incorporated in the formulations).
\par Our goal here is to minimize $\text{Var}({\hat{\theta}_{ML}})$ by considering the sensor gain vector $\ba$ as the optimization variable in both cases of centralized and decentralized estimation. In particular, the sensor gain optimization problem for both scenarios can be formulated as:
\begin{eqnarray}\label{eq:opt}
\max_{\ba} && {\ba}^{H}{\bH}^{H}\left({\bH \bD \bV  \bD^H \bH }^{H}+\bM\right)^{-1}{\bH \ba }  \\
\mbox{s. t.}
&& \ba \in \Theta,
\end{eqnarray}
where in the sequel, we denote the search space of the sensor vector $\ba$ by $\Theta$, and for a concise formulation we denote the covariance matrix of the observation noise by $\bV$ for both centralized and decentralized scenarios. It must be mentioned that the variance for the centralized and decentralized case follows the same mathematical structure, and although the structure of the channel matrix $\bH$ and the noise covariance matrices might be slightly different for the two scenarios, it does not affect the proposed formulation of the gain optimization framework in \eqref{eq:opt}. Briefly, the channel matrix used in the decentralized setting is a compressed version of the complete channel state information, and it only contains the CSI of the nodes chosen for diffusing information at each neighborhood. On the other hands, for the centralized scenario, the matrix $\bH$ contains the full CSI between the FC and the nodes.

\par Note that as $\bD=\mathbf{Diag}(\ba)$, the core matrix of the seemingly quadratic objective in (\ref{eq:opt}) depends on $\ba$. In the following, we will show that using a particular over-parametrization approach, the above  problem can be approached via a sequence of quadratic optimization problems.

Let $\eta=\eta_0 - {\ba}^{H}{\bH}^{H}\left({\bH \bD \bV  \bD^H \bH }^{H}+\bM\right)^{-1}{\bH \ba }$, and suppose that $\eta_0$ is sufficiently large to keep $\eta$ positive for all $\ba$.\footnote{To give an example of such $\eta_0$, as shown in the appendix, one can consider the following (although conservative) criterion to ensure the positivity of $\eta$:~ $\eta_0 > \frac{ N \, \| \bH \|_F^2}{\lambda_{\min}\left\{ \bM \right\} }$.} A detailed derivation of such an $\eta_0$ can be found in Appendix A. We seek to solve the following optimization problem:
\begin{eqnarray}\label{eq:opt-eta}
\min_{\ba} && \eta  \\
\mbox{s. t.}
&& \ba \in \Theta. \nonumber
\end{eqnarray}
Now let
\begin{equation}\label{eq:def1}
\bR \triangleq \left(
\begin{array}{ccc}
\eta_0 & | & \ba^H \bH^H \\
----& & ---------- \\
\bH \ba & | & \bH \bD \bV  \bD^H \bH^H+\bM
\end{array}\right),
\end{equation}
and note that $\be_1^H \bR^{-1} \be_1 = \eta^{-1}$ where $\be_1=(1~0~\cdots~0)^T$. In order to tackle (\ref{eq:opt-eta}), let $g(\by,\ba) \triangleq \by^H \bR \by$ (where $\by$ is an auxiliary vector variable), and consider the optimization problem:
\begin{eqnarray}\label{eq:opt-g}
\min_{\ba,\; \by} && g(\by,\ba)  \\
\mbox{s. t.} && \by^H \be_1= 1\quad \text{(or equivalently $y_1$ = 1),} \label{eq:const-y} \\
&& \ba \in \Theta. \label{eq:const-a}
\end{eqnarray}

The minimization of $g(\by,\ba)$ in \eqref{eq:opt-g} can be tackled via employing a \emph{cyclic optimization} approach with respect to $\ba$ and $\by$. Note that for fixed $\ba$, the minimizer $\by$ of (\ref{eq:opt-g}) is given by (see Result~35 in \cite[p. 354]{peterbook-spectral})
\begin{eqnarray} \label{eq:y_opt}
\by= \left( \frac{1}{\be_1^H \bR^{-1} \be_1 } \right) \, \bR^{-1} \be_1
\end{eqnarray}
which is the first column of $\bR^{-1} $ scaled in such a way to satisfy (\ref{eq:const-y}). A fast approach to computation of $\by$ in (\ref{eq:y_opt}) is as follows: Observe that $\by$ is a scaled version of the solution to the linear system  $\bR \by=\be_1$. Consequently, $\by$ is a scaled version of the vector orthogonal to all rows but the first row of $\bR$. Therefore, the direction of $\by$ can be easily obtained via the Gram-Schmidt process applied to the rows (excluding the first row) of $\bR$. Once the direction vector of $\by$ (i.e. $\by / \| \by \|_2$) is obtained, it can be scaled to achieve the optimal $\by$ in (\ref{eq:y_opt}) by simply making the first entry of $\by$ equal to one.

Interestingly, for a fixed $\by$, the minimization of  (\ref{eq:opt-g}) with respect to $\ba$ boils down to a quadratic optimization; see the following. We first note that a feasible $\by$ in (\ref{eq:opt-g}) can be partitioned as 
\begin{equation}
\by~\triangleq ~\left(
\begin{array}{c}
1\\
\widetilde{\by}
\end{array}\right)
\end{equation}
Then,
\begin{align}\label{eq:ga}
&~~~~\,\by^H \bR \by  \\
& = \left(
\begin{array}{c}
1\\
\widetilde{\by}
\end{array}\right)^H
\left(
\begin{array}{ccc}
\eta_0 & | & \ba^H \bH^H \\
---& & --------- \\
\bH \ba & | & \bH \bD \bV  \bD^H \bH^H+\bM
\end{array}\right)
\left(
\begin{array}{c}
1\\
\widetilde{\by}
\end{array}\right) \nonumber \\
& = C_1  + \nonumber \\
&~~~~
\left(
\begin{array}{c}
\ba \\
1
\end{array}\right)^H 
\underbrace{\left(
	\begin{array}{ccc}
	\left( \bH^H \widetilde{\by} \widetilde{\by}^H \bH \right) \odot \bV & | & \bH^H \widetilde{\by}  \\
	---------& & --- \\
	\widetilde{\by}^H \bH  & | & 0
	\end{array}\right)}_{\triangleq \bQ}
\left(
\begin{array}{c}
\ba \\
1
\end{array}\right) \nonumber
\end{align}
where $C_1 = \eta_0 + \widetilde{\by}^H \bM \widetilde{\by}$ is invariant with respect to the sensor gain vector $\ba$, and we have used the identity
\begin{eqnarray}
\widetilde{\by}^H \bH \bD \bV \bD^H \bH^H \widetilde{\by} \nonumber = \ba^H \left( \left( \bH^H \widetilde{\by} \widetilde{\by}^H \bH \right) \odot \bV \right) \ba.
\end{eqnarray}
As a result, the minimization of (\ref{eq:opt-g}) with respect to $\ba$ can be recast as
\begin{eqnarray}\label{eq:opt-Q}
\min_{\ba} &&  \left(
\begin{array}{c}
\ba \\
1
\end{array}\right)^H  \bQ ~\left(
\begin{array}{c}
\ba \\
1
\end{array}\right)  \\ 
\mbox{s. t.}
&& \ba \in \Theta\;. \nonumber
\end{eqnarray}
Note that when $\Theta$ enforces a fixed-energy constraint for $\ba$ (i.e. $\| \ba \|_2^2=N$), (\ref{eq:opt-Q}) is equivalent to
\begin{eqnarray}\label{eq:opt-Qtilde}
\max_{\ba} &&  \left(
\begin{array}{c}
\ba \\
1
\end{array}\right)^H  \widetilde{\bQ} ~\left(
\begin{array}{c}
\ba \\
1
\end{array}\right)  \\ 
\mbox{s. t.}
&& \ba \in \Theta \nonumber
\end{eqnarray}
where $\widetilde{\bQ} \triangleq \lambda \bI_{N+1} - \bQ$, and $\lambda > \lambda_{\max} (\bQ)$. 
Using the following power-method-like iterations, the objective of (\ref{eq:opt-Qtilde}) can be made to be monotonically increasing, and the objective of (\ref{eq:opt-g}) monotonically decreasing (see \cite{soltanalian2013designing}\nocite{soltanalian2013joint}-\cite{soltanalian2014single} for details):
\begin{eqnarray}\label{eq:power}
\min_{\ba^{(t+1)}} &&  \left\|  \left(
\begin{array}{c}
\ba^{(t+1)} \\
1
\end{array}\right)^H  - \, \widetilde{\bQ} \, \left(
\begin{array}{c}
\ba^{(t)}  \\
1
\end{array}\right) \right\|_2  \\ 
\mbox{s. t.}
&& \ba \in \Theta, \nonumber
\end{eqnarray}
where $t$ denotes the iteration number, and $\ba^{(t)}$ is the current value of $\ba$.

By calculating (\ref{eq:y_opt}), it is now straightforward to verify that at the minimizer $\by$ of (\ref{eq:opt-g}),
\begin{eqnarray}
g(\by,\ba)=\eta.
\end{eqnarray}
Therefore, each step of the cyclic optimization of (\ref{eq:opt-g}) with respect to $\by$ and $\ba$ leads to a decrease of $\eta$. More concretely, observe that if $f(\ba)=\eta$ then
\begin{eqnarray}\label{eq:converge}
f\left(\ba^{(k+1)}\right) &=& f\left(\by^{(k+2)},\ba^{(k+1)}\right) \\ \nonumber
&\leq & f\left(\by^{(k+1)},\ba^{(k+1)}\right) \\\nonumber
&\leq & f\left(\by^{(k+1)},\ba^{(k)}\right)\quad = \, f\left(\ba^{(k)}\right)
\end{eqnarray}
where the index $k$ denotes the iteration number.
\vspace{5pt}

\emph{Remark 2:}
When $\Theta$ represents the finite-energy constraint, problem (\ref{eq:opt-g}) is biconvex in $(\by,\ba)$, and (\ref{eq:opt-Qtilde}) is simply a quadratically constrained quadratic program (QCQP). In particular, our cyclic approach described above boils down to an \emph{alternate convex search (ACS)}; see \cite[p. 393]{Gorski2007} for details. \hfill $\blacksquare$

In the following, we consider various practical constraints on the gain vector $\ba$ and provide the corresponding power method-like iterations for each case.
\subsection{Practical Signal Constraints}
\par In a practical model,  the sensor gains should always be bounded by a finite-energy constraint. However, the variance expression in (\ref{eq:var}) is a monotonically decreasing function of the energy of $\ba$ (i.e. $\| \ba \|^2_2$) which implies that (\ref{eq:var}) attains its minimum only if the sensor network employs the maximum energy possible. In light of the latter observation, we consider a number of possible sensor signal constraints, including:

\begin{itemize}
	\item[(a)] Finite or fixed energy:
	\begin{eqnarray}
	\|\ba\|_2^2=N.
	\end{eqnarray}
	\item[(b)] Phase-shift only:
	\begin{eqnarray}
	|a_i|=1,\quad i \in \{ 1,\cdots, N \}.
	\end{eqnarray}
	\item[(c)] Phase-shift only with quantized phase values:
	\begin{eqnarray}
	a_i \in \left\{ e^{j\frac{2 \pi}{Q} q}: ~q  \in \integerZ_Q \right\},\quad i \in \{ 1,\cdots, N \}.
	\end{eqnarray}

	\item[(d)] Sensor selection: Only $K<N$ of the sensors can transmit, viz.
	$		\|\ba\|_0 \leq K$
	which may be combined with a finite-energy constraint, i.e. $	\|\ba\|_2^2=N$, or the phase-shift only constraint at the non-zero entries of $\ba$.
\end{itemize}

In the following subsection, we provide the corresponding power method-like iterations for each constraint---more on this below.
\nocite{akyildiz2002survey,mudumbai2007feasibility}
\nocite{stoica2004cyclic}

\subsection{Constrained Solutions to (\ref{eq:power})}
Let $\widehat{\ba}^{(t)}$ denote the vector comprising the first $N$ entries of $\widetilde{\bQ} \, \left(
\begin{array}{c}
\ba^{(t)}  \\
1
\end{array}\right)$, viz.
$
\widehat{\ba}^{(t)} = (\bI_N~\bzero_{N \times 1})  ~ \widetilde{\bQ} \, \left(
\begin{array}{c}
\ba^{(t)}  \\
1
\end{array}\right).
$ 
The solutions to (\ref{eq:power}) for different sensor gain constraints $\Theta$ are given by:
\begin{itemize}
	\item[(a)]  \underline{Finite or fixed energy}:
	\begin{eqnarray}
	\ba^{(t+1)} = \left( \sqrt{N} / \| \widehat{\ba}^{(t)}   \|_2 \right) \, \widehat{\ba}^{(t)}  .
	\end{eqnarray}

	\item[(b)] \underline{Phase-shift only}:
	\begin{eqnarray}
	\ba^{(t+1)} = \exp \left( j \arg \left(\widehat{\ba}^{(t)}  \right) \right).
	\end{eqnarray}
	
		\item[(c)] \underline{Phase-shift only with quantized phase values}:
		\begin{eqnarray}
		\ba^{(t+1)} = \exp \left( j \mu_Q \left( \arg \left(\widehat{\ba}^{(t)}  \right)\right) \right)		\end{eqnarray}

where $\mu_Q(.)$ yields (for each entry of the vector argument) the closest element in the $Q$-ary alphabet.
\vspace{3pt}

	\item[(d)] \underline{Sensor selection}: Note that for the optimal $\ba^{(t+1)}$ of (\ref{eq:power}) we have that
	$\arg(\ba^{(t+1)})=\arg(\widehat{\ba}^{(t)})$. Therefore, we can exclude the phase variables while finding the absolute values of the entries of the optimal $\ba^{(t+1)}$. In fact, without loss of generality, we can assume that both $\ba^{(t+1)}$ and $\widehat{\ba}^{(t)}$ are real-valued and non-negative. Note that
	\begin{eqnarray}
	 \left\| \ba^{(t+1)}- \widehat{\ba}^{(t)} \right\|_2^2 = C_2 - 2 \, \ba^{T\, (t+1)} \widehat{\ba}^{(t)},
	\end{eqnarray}
	where $C_2 = \| \ba^{(t+1)} \|_2^2 +\| \widehat{\ba}^{(t)}\|_2^2 $ is constant. According to a theorem due to Hardy, Littlewood, and Polya \cite{day1972rearrangement}, the inner product of $\ba^{(t+1)}$ and $\widehat{\ba}^{(t)}$ can be maximal only if the elements of $\ba^{(t+1)}$ are sorted to have the same order of \emph{magnitude} as in $\widehat{\ba}^{(t)}$. Consider the $K$ elements in $\widehat{\ba}^{(t)}$ with maximum absolute values, and let $\bs$ be a binary ($0/1$) vector that is one only in the corresponding locations of these largest $K$ elements, and is zero otherwise. Then the optimal $\ba^{(t+1)}$  of (\ref{eq:power}) becomes 
	\begin{eqnarray}
	\ba^{(t+1)} = \sqrt{N} \, \left( \frac{\widehat{\ba}^{(t)} \odot \bs }{\left\|\widehat{\ba}^{(t)} \odot \bs\right\|_2} \right) .
	\end{eqnarray}
	Similarly, if the non-zero entries of $\ba^{(t+1)}$ are to be constant-modulus (i.e., phase-shift only and sensor selection scenario) then the optimal $\ba^{(t+1)}$  can be obtained as
		\begin{eqnarray}
		\ba^{(t+1)} = \sqrt{N/K} \, \left( \exp \left(j \arg\left(\widehat{\ba}^{(t)} \right)  \right) \odot \bs \right) .
		\end{eqnarray}

\end{itemize}	
\vspace{4pt}
	Finally, the proposed method is summarized in Table~\ref{table:method}.

	\begin{table}[tp]\label{tb:one}
		\footnotesize
		\caption{The Proposed Sensor Gain Optimization Approach} \label{table:method} \centering
		\begin{tabular}{p{3.3in}}
			\hline \hline
			\textbf{Step 0}: Initialize the auxiliary vector $\by$ with a random vector in $\complexC^{N+1}$ such that $y_1=1$. Initialize $\ba \in \Theta$.\\
			\textbf{Step 1}: Employ the quadratic formulation in (\ref{eq:opt-Qtilde}), and particularly the power method-like iterations in  (\ref{eq:power}) to update the sensor gain vector $\ba$ (until convergence).\\
			\textbf{Step 2}: Update $\by$ using (\ref{eq:y_opt}), or by employing the fast approach discussed below~(\ref{eq:y_opt}).\\
			\textbf{Step 3}: Repeat steps 1 and 2 until a pre-defined stop criterion  is satisfied, e.g. $\left|f(\ba^{(k)})-f(\ba^{(k+1)})\right| \leq \xi$ for some $\xi>0$, where $k$ denotes the outer-loop iteration number, and $f(\cdot)$ is the criterion to be optimized (e.g., $f(\mbf{a})= \eta$).\\
			\hline \hline
		\end{tabular}
	\end{table}
	
\begin{figure*}[t]
	\begin{minipage}[b]{0.48\linewidth}
		\centering
		\centerline{\includegraphics[width=7.1cm]{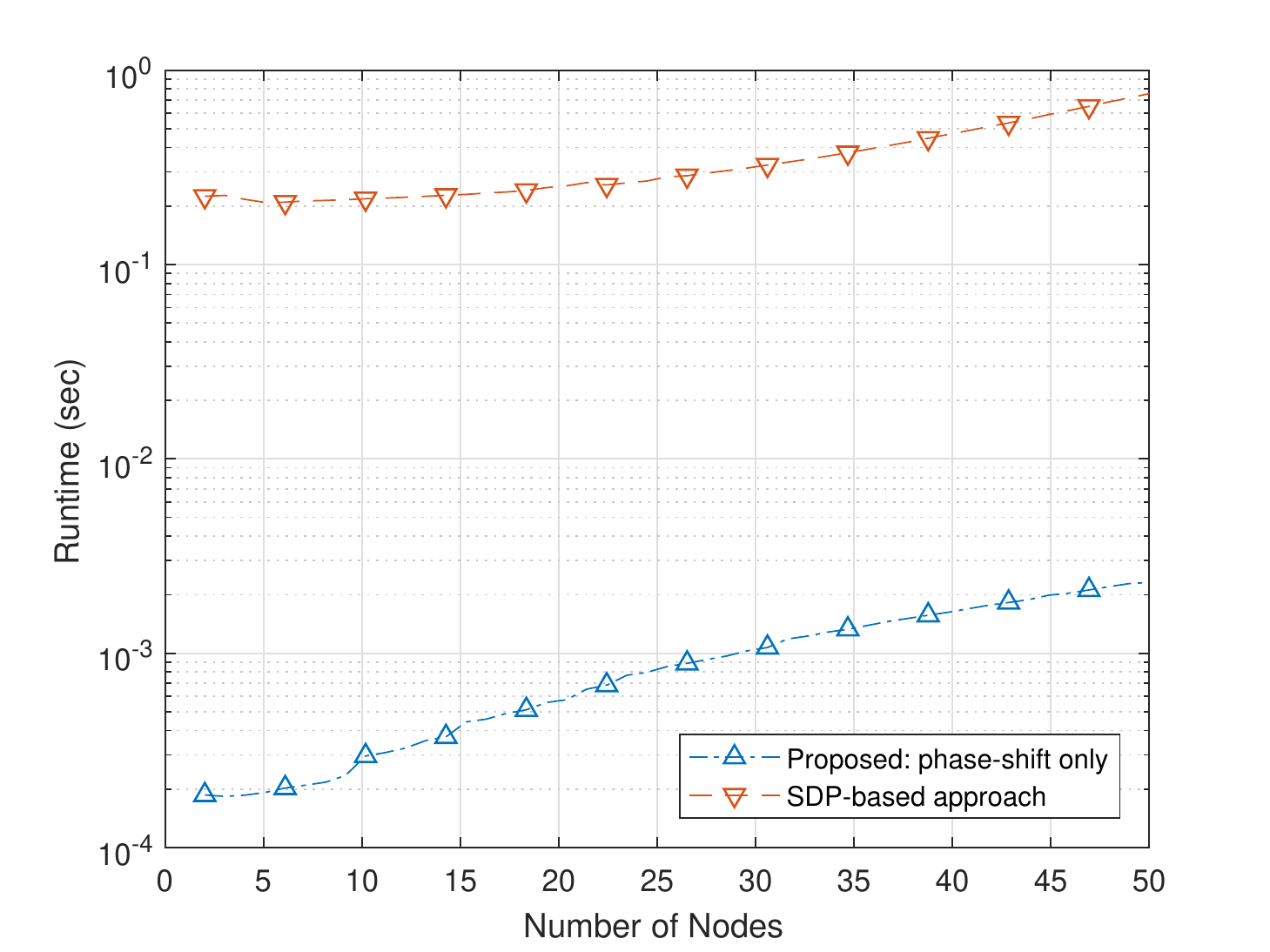}}
		\centerline{(a)}\medskip
	\end{minipage}
	\hfill
	\begin{minipage}[b]{0.48\linewidth}
		\centering
		\centerline{\includegraphics[width=7.1cm]{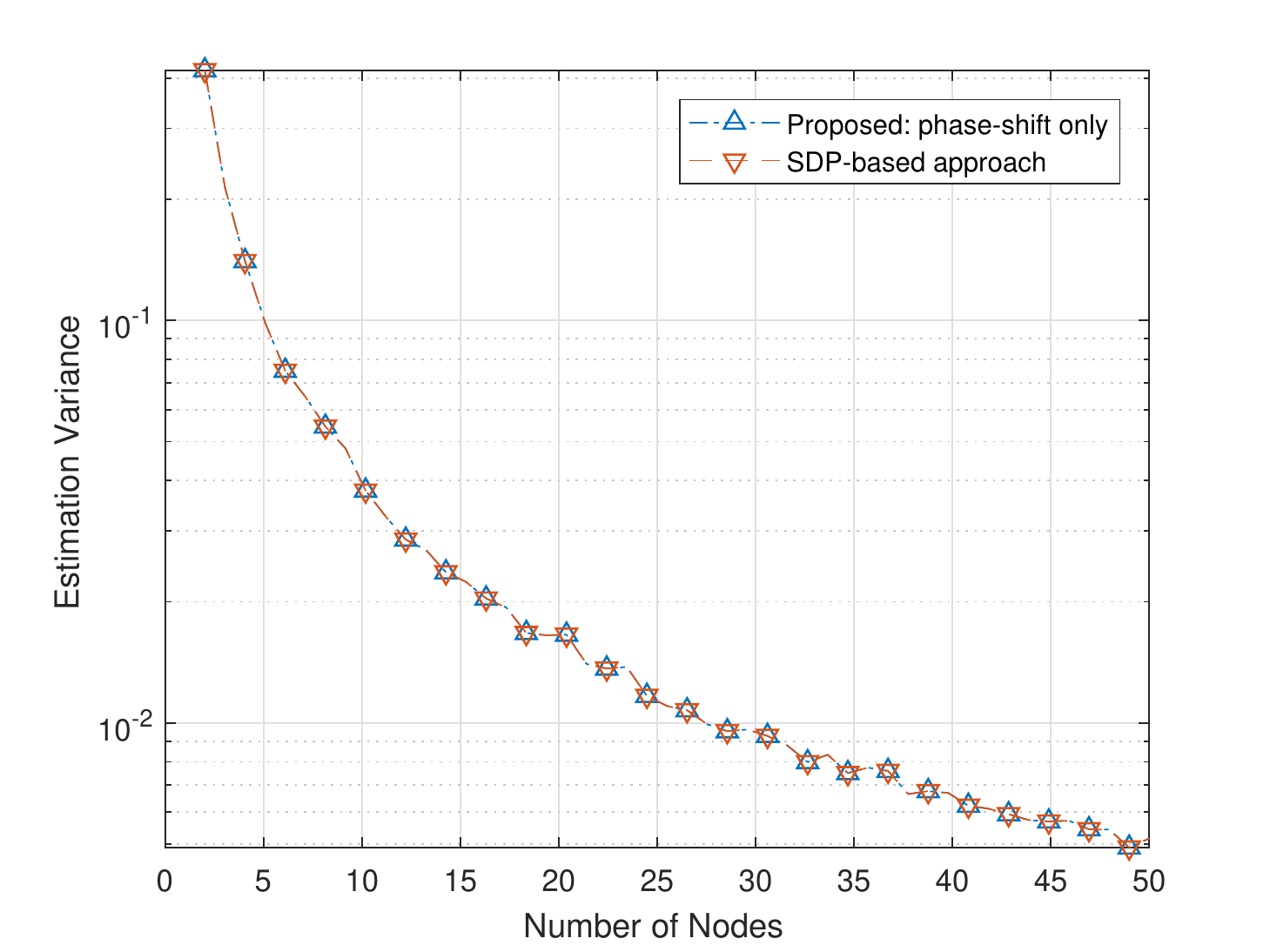}}
		\centerline{(b)}\medskip
	\end{minipage}
	\caption{Comparison of (a) the runtime and (b) the estimation variance of the proposed method and the SDP-based approach of \cite{jiang2013estimation}, for the decentralized scenario. The proposed algorithm exhibits significantly lower computational cost, while achieving a similar estimation variance.}
	\label{fig:dec1}
\end{figure*}
\begin{figure}[t]
	\begin{minipage}[b]{\linewidth}
		\centering
		\centerline{\includegraphics[width=7.1cm]{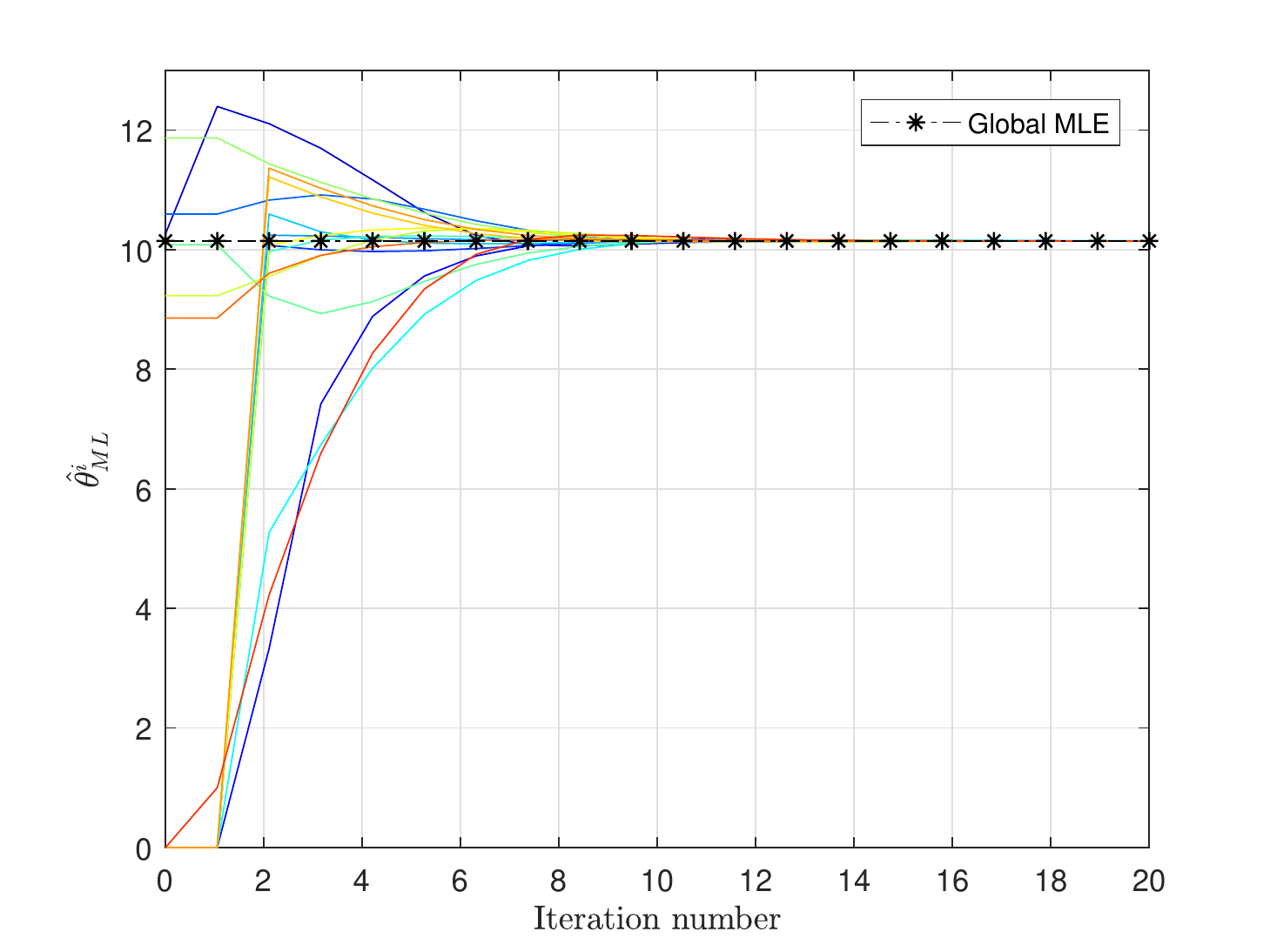}}
	\end{minipage}
	\caption{Convergence of the ADMM-based ML estimate.}
	\label{fig:dec2}
\end{figure}
		
\subsection{Phase-Shift Only Case---A Simplified Approach} \label{subsec:phaseshift}
We note that, in phase-shift only scenarios, the problem in (\ref{eq:opt}) becomes a unimodular quadratic program (UQP) \cite{soltanalian2013designing} \nocite{zhang-complexquad} and we can deal with it more easily compared to the general case. Namely, (\ref{eq:opt}) can be rewritten as
\begin{eqnarray}\label{eq:opt-phaseshift}
\max_{\ba} && \ba^H \bB \ba  \\
\mbox{s. t.}
&& |a_i|=1,\quad i \in \{ 1,\cdots, N \}, \nonumber
\end{eqnarray}
in which $\bB$ is not dependent on $\ba$. The power method-like iterations for  (\ref{eq:opt-phaseshift}) are simply given by
\begin{eqnarray}
\ba^{(k+1)} = \exp \left( j \arg \left( \bB \ba^{(k)} \right) \right)
\end{eqnarray}
and yield a monotonically increasing objective function in (\ref{eq:opt-phaseshift}). We refer the interested reader to find more details on the properties of power method-like iterations in \cite{soltanalian2013designing}\nocite{soltanalian2013joint}-\cite{soltanalian2014single}.

\emph{Remark 3:}
A brief computational analysis of the proposed method (see Table I) is as follows. Employing the power method-like iterations for updating the sensor gain vector $\ba$ has a complexity of $\mathcal{O}(LN^2)$ where $L$ denotes the total number of iterations that are performed within each outer-loop iteration of the proposed optimization method. On the other hand, the design of $\by$ using \eqref{eq:y_opt} has a $\mathcal{O}(M^{2.38})$ complexity per-iteration while the proposed fast approach for computation of $\by$ described below $\eqref{eq:y_opt}$ has a complexity of $\mathcal{O}(M^2)$. Hence, the optimization method (summarized in Table I) has a complexity of $\mathcal{O}\left(\mbox{max}\{LN^2, M^2\}\right)$ per-iteration (if the proposed fast method is used for finding $\by$ at each iteration). Further note that the proposed algorithm yields a monotonically decreasing objective function according to \eqref{eq:converge}.\hfill $\blacksquare$	
\section{Numerical Results}	 \label{sec:num}
To evaluate the performance of the proposed optimization framework, in this section, we present several numerical examples for both centralized and decentralized estimation scenarios. For a fair comparison, we set the parameters to the same values as those in \cite{jiang2013estimation}: The sensor nodes are configured with a single antenna, and the FC is assumed to have four antennas. The wireless fading channel coefficient between sensor node $i$ and the FC is modeled as $\mathbf{h}_i=e^{j\gamma_i} / d_i^\alpha$, where the phase argument $\gamma_i$ is uniformly distributed over $[0,2\pi)$, $d_i$ is the distance between the sensor node and FC, and the path loss exponent $\alpha$ is set to $1$. In all plots, the results are
obtained by averaging the outcomes over $300$ random channel realizations. 
\begin{figure}[t]
	\centering
	
	\includegraphics[width=6.9cm]{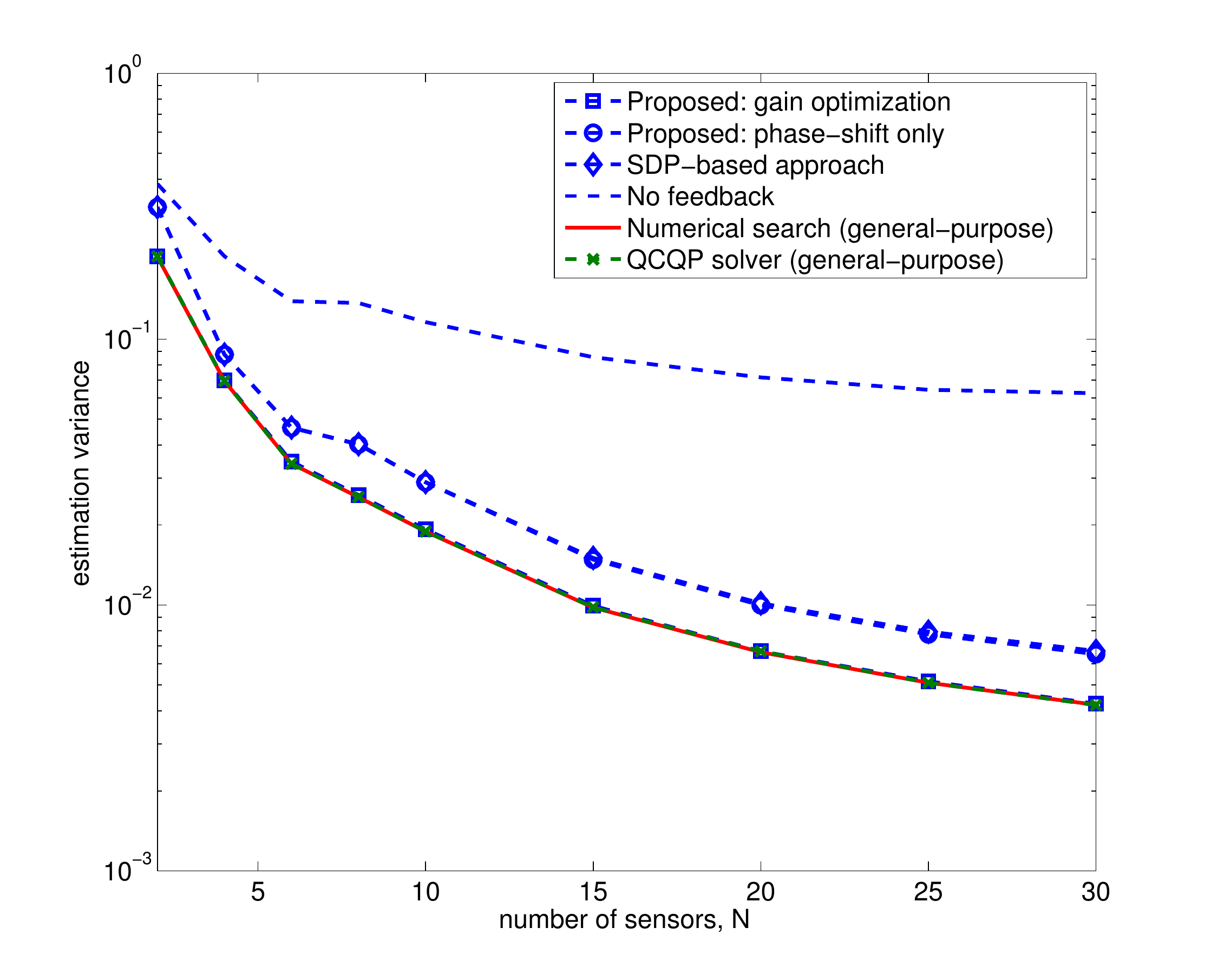}
	\caption{Estimation performance of various sensor gain/phase-shift optimization approaches vs. number of sensors. The comparison is performed with the SDP-based approach of \cite{jiang2013estimation}, the no-feedback case (with $\ba=\bone$), a numerical search approach, as well as by employing a general-purpose QCQP solver for tackling (\ref{eq:opt-Qtilde}) in lieu of the power method-like iterations.}\label{f1}

	\includegraphics[width=6.9cm]{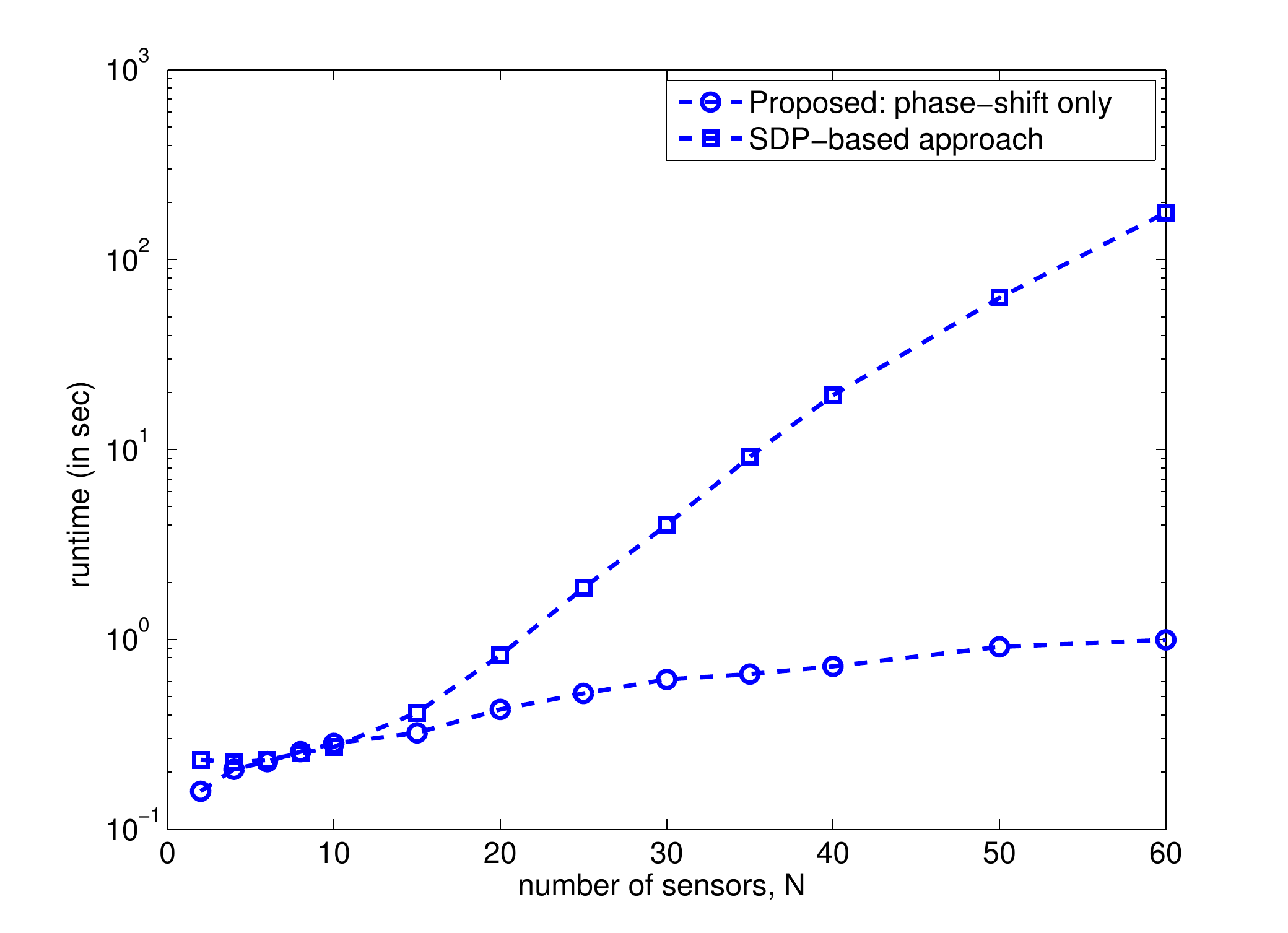}
	\caption{Comparison of runtime for the proposed method and the state-of-the-art SDP-based approach of \cite{jiang2013estimation}. The proposed method exhibits a significantly lower computational cost compared to the SDP-based approach, particularly when $N$ grows large.}\label{f2}

	\includegraphics[width=6.9cm]{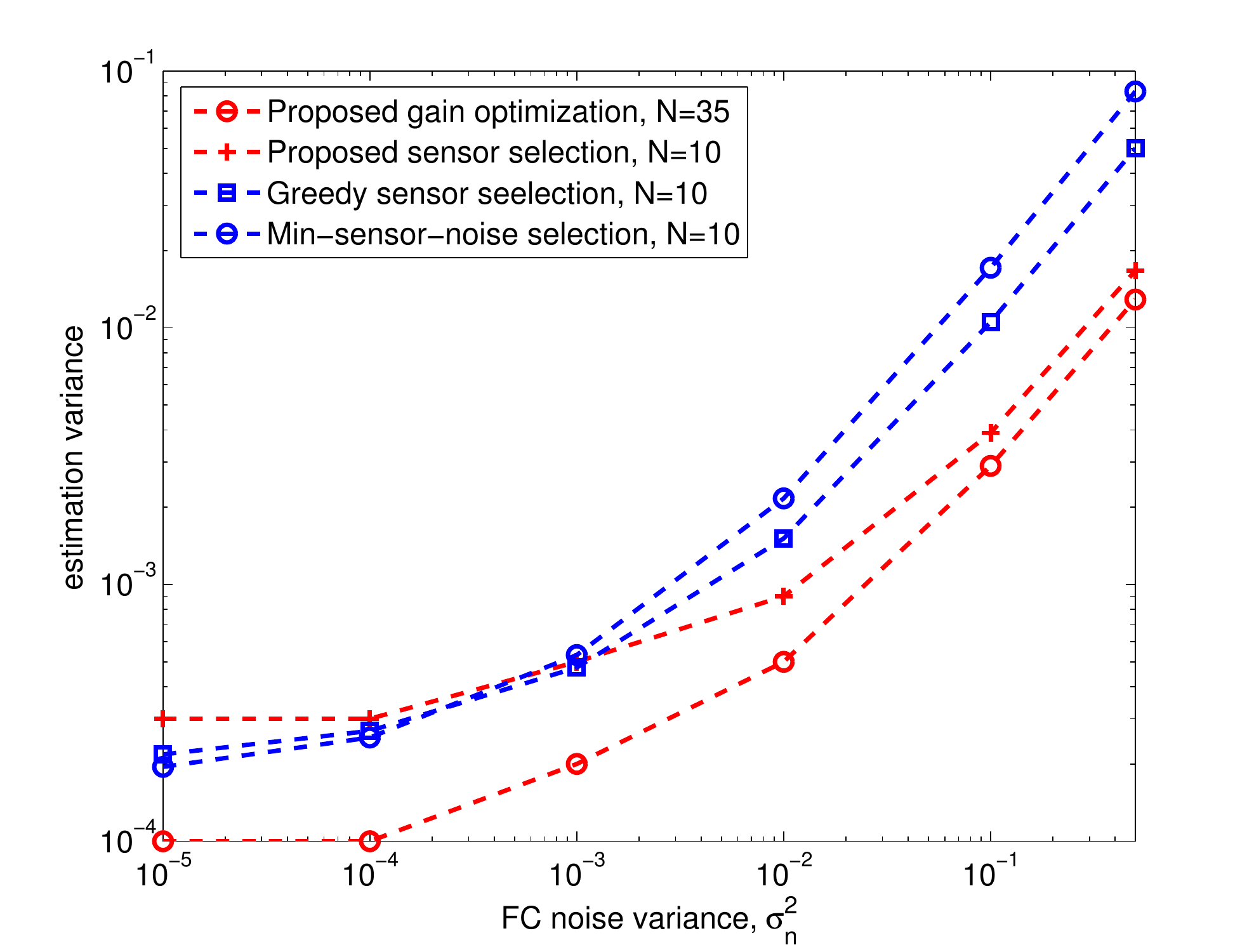}
	\caption{Performance of various sensor selection algorithms ($10$ sensor nodes are selected out of $35$ sensors). The greedy and min-sensor-noise selection methods were proposed in \cite{jiang2013estimation}.}\label{f3}
\end{figure}

\subsection{Decentralized Parameter Estimation Scenario}

\label{sec:typestyle}
In this part, we investigate the performance of our proposed sensor gain optimization algorithm for the decentralized and distributed parameter estimation scenario. 
We compare our sensor gain optimization algorithm (Table I) with the SDP-based approach of \cite{jiang2013estimation}.

Fig. \ref{fig:dec1}(a) shows a comparison of the computational cost (machine runtime)  between our algorithm and the SDP-based approach in \cite{jiang2013estimation}. It is observed from Fig. \ref{fig:dec1}(a) and Fig. \ref{fig:dec1}(b) that although the two algorithms yield similar estimation variance, our proposed optimization algorithm has a significantly lower computational burden. For example, with $N=50$ nodes, the runtime of our algorithm is less than $1\%$ of the runtime associated with the SDP-based approach. This is particularly of importance in WSNs since not only are the processing resources of the nodes limited, the environmental parameters (e.g., the channels) might change and need frequent reassessment. Hence, it is important for the network to be able to adapt to the new environment \textit{as quickly as possible} with \textit{minimal cost}.

Our proposed two-stage algorithm also enables the nodes to obtain the global ML estimate of the parameter based on their \textit{local} information by applying the distributed fusion algorithm described in subsection~2.A. Fig. \ref{fig:dec2} illustrates the simulation results for this ADMM-based decentralized estimation and the convergence of the proposed decentralized MLE algorithm to that of the global MLE for a network with $N=16$, and $\theta=10$. We see that the local estimate of each node $\hat{\theta}^i_{ML}(k)$ converges to the global MLE of the parameter computed in \eqref{eqs:13}, and a consensus is thus achieved quickly.

\subsection{Centralized Parameter Estimation Scenario}
We begin by comparing the estimation variance of different sensor phase (or gain) optimization methods, namely (i) the proposed gain optimization method (Table~\ref{table:method}) with finite (or fixed energy) constraint, (ii) the proposed phase-shift only approach (subsection \ref{subsec:phaseshift}), (iii) the phase-shift only solutions provided by the semidefinite programming (SDP) approach of \cite{jiang2013estimation}, and (iv) the no feedback case. For the no feedback case, we assume that there is no feedback channel between the FC and sensor nodes and that the vector $\ba$ is set to a vector of all ones. 
The results are shown in Fig.~\ref{f1}.  We see that after phase or gain optimization the estimation error is considerably reduced \emph{($\sim$ by a factor of $10$)} compared to the no feedback case. Moreover, our proposed phase-shift only approach can achieve an estimation variance almost identical to that of the SDP-based approach of \cite{jiang2013estimation}. Although our proposed method and the SDP-based approach exhibit the same performance in this scenario, we will later show that the proposed power-method iterations have a significantly lower computational burden compared to that of the SDP-based approach. To assess the quality of our design in the complex gain optimization case, we also resort to comparison with (i) a general-purpose numerical search algorithm based on the \emph{active set} method \cite{wright1999numerical}, as well as (ii) solving the optimization problem in (\ref{eq:opt-Qtilde}) by a general-purpose QCQP solver in lieu of directly employing the power method-like iterations. In terms of the estimation variance, the performance of the proposed gain optimization appears to be identical to the numerical search. Additionally, the low-complexity power method-like iterations can achieve the same performance as the QCQP solver, which verifies the optimality of the results obtained by power method-like iterations---as expected due to the convexity of (\ref{eq:opt-Qtilde}) in the finite-energy scenarios.

A comparison of the computational cost (machine runtime) between the state-of-the-art SDP-based approach of \cite{jiang2013estimation} and the proposed algorithm is presented in Fig.~\ref{f2} (the results were obtained on a standard PC with a 2.40GHz CPU and 3.5 GB memory). Due to the fact that the SDP-based approach can only handle the phase-shift case, we consider a comparison with the phase shift-only method described in subsection \ref{subsec:phaseshift}. As shown earlier in Fig.~\ref{f1}, the two methods yield similar estimation variance. However, according to Fig.~\ref{f2}, the computational cost of the proposed phase shift-only approach is significantly smaller compared to the SDP-based method. More precisely, while for a small number of sensors ($N \leq 10$)
the two methods have similar cost, the advantage of the proposed approach becomes more clear when $N$ grows large. For instance,  with $N=60$ sensors, the runtime of the proposed approach is less than $ 1 \%$ of that associated with the SDP-based method.

Finally, Fig.~\ref{f3} investigates the performance of the sensor selection problem (with a fixed-norm sensor gain) for a scenario in which $10$ sensors are to be selected out of $35$ sensor nodes for signal transmission. We set the covariance matrix of the noise at the FC to $\bM= \sigma_n^2 \bI$ with variable noise variance $\sigma_n^2$. The simulation results show that when the additive noise at the FC is small, the performance of the proposed sensor selection method is close to the greedy or min-sensor-noise methods devised in \cite{jiang2013estimation}, whereas, when the additive noise grows large, its performance is superior to that of both the greedy and min-sensor-noise methods. It is interesting to observe that, for large values of the FC noise variance, the performance of the proposed sensor selection algorithm is relatively close to the case when all the $35$ sensor nodes are used for signal transmission.

%
%
%
%
%

\section{Conclusions} \label{sec:conclusion}
In this work, we considered the problem of transmission gain optimization in a distributed wireless sensor network for both centralized and decentralized parameter estimation scenarios. We proposed an efficient sensor gain optimization framework which enables us to effectively reduce the parameter estimation variance resulting in a far better estimation accuracy in both the centralized and decentralized cases. The proposed optimization framework is based on the power method-like iterations and can deal with the optimization of complex gains of the sensors, and furthermore, can handle various sensor gain constraints including e.g., finite or fixed energy, phase-shift only (with a quantized phase values as a simple extension) and sensor selection cases, and was shown to exhibit a superior performance in large-scale sensor networks. In addition, the suggested framework enjoys a low computational cost compared to the SDP-based approach, and thus, can be a good candidate for large-scale sensor networks that may need adaptive sensor gain optimization in real-time. Moreover, we extended our sensor gain optimization algorithm to the decentralized parameter estimation scenario in which a consensus-based algorithm in conjunction with graph signal processing methods are employed to perform the parameter estimation with optimized transmission gains in a decentralized manner.\looseness=-1

\appendices	
\section{Selection of $\eta_0$}\label{sec:eta0}\label{app:one}
In order to ensure that $\eta$ stays positive, we choose $\eta_0$ such that
\begin{equation}
\eta_0 > N \, \lambda_{\max} \left\{  {\bH}^{H}\left({\bH \bD \bV  \bD^H \bH }^{H}+\bM\right)^{-1}{\bH }\right\}.
\end{equation}
Note that
\begin{eqnarray}
&& \lambda_{\max} \left\{  {\bH}^{H}\left({\bH \bD \bV  \bD^H \bH }^{H}+\bM\right)^{-1}{\bH }\right\} \\ \nonumber &\leq& \tr \left\{  {\bH}^{H}\left({\bH \bD \bV  \bD^H \bH }^{H}+\bM\right)^{-1}{\bH  }\right\}
\\ \nonumber &\leq& \frac{\tr\left\{\bH^H \bH\right\}}{\lambda_{\min}\left\{ \bM \right\} }~=~ \frac{ \| \bH \|_F^2}{\lambda_{\min}\left\{ \bM \right\} }.
\end{eqnarray}
As a result, it would be sufficient if we choose
\begin{eqnarray}
\eta_0 > \frac{ N \, \| \bH \|_F^2}{\lambda_{\min}\left\{ \bM \right\} }.
\end{eqnarray}

\bibliographystyle{IEEEtran}
\balance
\bibliography{MS_refs,refs,strings,IEEEabrv_Asilomar1,IEEEabrv_Asilomar2}
\end{document}